\documentclass[aps,prl,twocolumn,groupedaddress,showpacs,floatfix]{revtex4}
\usepackage{graphicx}
\usepackage{epstopdf}
\usepackage{amssymb}

\begin{document}

\title{Self-driven nonlinear dynamics in magneto-optical traps}

\author{T.\  Pohl}
\affiliation{ITAMP, Harvard-Smithsonian Center for Astrophysics, 60 Garden Street, Cambridge MA 02138}
\author{G.\ Labeyrie}
\author{R.\ Kaiser}
\affiliation{Institut Non Lin\'eaire de Nice, UMR 6618, 1361 route des Lucioles, F-06560, Valbonne}

\date{\today}

\begin{abstract}
We present a theoretical model describing recently observed collective effects in large magneto-optically trapped atomic ensembles. Based on a kinetic description we develop an efficient test particle method, which in addition to the single atom light pressure accounts for other relevant effects such as laser attenuation and forces due to multiply scattered light with position dependent absorption cross sections. Our calculations confirm the existence of a dynamical instability and provide deeper insights into the observed system dynamics.
\end{abstract}


\maketitle

Since its first realization in 1987 \cite{Raa87} the magneto-optical trap (MOT) has become a standard technique for providing a robust source of large numbers of cold atoms. While multiple scattering of the absorbed laser-light is known as a major limitation for achieving Bose-Eintein condensation, it also leads to interesting collective effects which have been  studied over the last years \cite{Wal90,Ses91,Ste92,Bag93,Ara95,Tow95} and a variety of static structures has been observed and investigated  by different theoretical approaches \cite{Ses91,Bag93,Pru00,Oli04}.

Only recently, experiments have revealed a so far unexplored dynamical instability in 
three-dimensional MOTs connected with the appearance of self-excited radial oscillations 
\cite{Lab05}, which constitutes a complex nonlinear dynamics phenomenon. Understanding 
the observed effect turns out to be of broader interest, as it provides a 
clean laboratory realization of similar plasma- and astrophysical phenomena, such 
as, e.g., pulsating stars \cite{Cox80}, which are generally difficult to access.

Here we develop a theoretical model, describing the observed instability and providing a 
physical picture of the underlying mechanism.
As discussed in \cite{Tow95}, sub-Doppler cooling mechanisms only affect a very small 
fraction of large trapped atom clouds. 
Hence, the overall behavior of large atomic ensembles is well described within a basic 
Doppler-cooling picture, where the resulting trapping force along each laser beam can be 
written as \cite{Let89,Met99} 
\begin{equation} \label{e1}
F_{{\rm trap}}^{(i)}(x,v)=\frac{\hbar \Gamma}{2}\left(s_{+}\sigma_{+}(x,v)-s_{-}\sigma_{-}(r,v)\right) \:,
\end{equation}
where
\begin{equation} \label{e2}
\sigma_{\pm}=\sigma_0\left(1+3(s_{+}+s_{-})+4\frac{(\delta\mp kv\mp \mu x)^2}{\Gamma^2}\right)^{-1}
\end{equation}
is the absorption cross section for the two laser beams (including a saturation by the 3 pairs of laser beams), $\sigma_0=3\lambda/2\pi$ the on-resonance 
absorption cross 
section, $\lambda$ the laser wavelength, $\Gamma$ the transition linewidth, $\delta$ the detuning from 
resonance, $\mu x$ determines 
the Zeeman shift of the atomic transition due to the MOT 
magnetic field and $s_{\pm}=I_{\pm}/I_{\rm sat}$ denotes the saturation 
parameter of the respective laser beam of intensity $I_{\pm}$ with $I_{\rm sat}$ being the saturation 
intensity of the atomic transition. For the discussion below it is convenient to split the force 
according to $F_{{\rm trap}}(x,v)=\alpha(|x|)x+\beta(|x|,|v|)v$, with $\alpha(|x|)=F_{{\rm trap}}(|x|,0)/|x|$ 
and $\beta(|x|,|v|)=(F_{{\rm trap}}(|x|,|v|)-F_{{\rm trap}}(|x|,0))/|v|$.

In order to simplify our theoretical considerations we use the following spherical symmetric generalization of eq.(\ref{e1})
\begin{equation} \label{e3}
{\bf F}_{\rm trap}=\alpha(r){\bf r}+\beta(r,v){\bf v}\;,
\end{equation}
While experimental confinement configurations generally do not obey this symmetry, eq.(\ref{e3}) describes the important features of the resulting force in both the linear and nonlinear trapping regions. 

At higher densities, attenuation of the laser light inside the cloud, results in an additional effective confining force experienced by the atoms \cite{Dal88}. To account for this effect within our spherical symmetry assumption, the spatial intensity profile is obtained from
\begin{eqnarray} \label{e4}
s_+&=&s_0e^{-\int_r^{\infty}\sigma_{+}(r^{\prime})\rho(r^{\prime})}\;,\nonumber\\
s_-&=&s_0e^{-\int_0^{\infty}\sigma_{+}(r^{\prime})\rho(r^{\prime})-\int_0^{r}\sigma_{-}(r^{\prime})\rho(r^{\prime})}
\end{eqnarray}
where $s_0$ is the saturation parameter of the incident beam.
Moreover, multiple scattering of the absorbed laser light inside the cloud leads to an additional outward directed pressure, 
caused by an effective interaction between the atoms \cite{Wal90}. Neglecting higher order scattering events, which are known to screen the atom-atom interaction \cite{Ell94}, a photon scattered off an atom at position ${\bf r}_1$ exerts an 
average force on an absorbing atom at ${\bf r}_2$ according to \cite{Wal90,Ste92}
\begin{equation} \label{e5}
{\bf F}_{\rm rsc}=\frac{3I_{\rm sat}}{4\pi c}\left(s_+\sigma_{+}\sigma_{\rm rsc}^{(+)}+s_-\sigma_{-}\sigma_{\rm rsc}^{(-)}\right)\frac{{\bf r}_2-{\bf r}_1}{\left|{\bf r}_2-{\bf r}_1\right|^3}\;.
\end{equation}
The reabsorption cross section $\sigma_{\rm rsc}^{(+/-)}$ is obtained by convolving the absorption cross section of the emitted light with the emission spectrum of the atom at ${\rm r}_1$ in the presence of either left or right circularly polarized laser light. Note that $\sigma_{\rm rsc}$ may depend on both coordinates via the space dependence of the local laser intensities as well as of the respective detunings. 
Previously \cite{Ses91,Ste92,Tow95,Pru00,Oli04,Arn00}, such coordinate 
dependencies have been neglected, which according to eq.(\ref{e5}) 
results in a Coulomb-like interaction with effective charges, again 
underlining the close analogy with plasma and gravitational physics 
problems. 
In large clouds, however, we find the position dependence 
of the effective charges to be important for the static and dynamics 
properties of the trapped atom cloud.

Starting from eqs.(\ref{e2})-(\ref{e5}) the collective system dynamics is described by the following kinetic equation
\begin{widetext}
\begin{equation} \label{e6}
\frac{\partial f}{\partial t}+{\bf v}\frac{\partial f}{\partial {\bf r}}+M^{-1}\alpha(r){\bf r}\frac{\partial f}{\partial {\bf v}}+M^{-1}{\bf F}_{\rm mf}({\bf r})\frac{\partial f}{\partial {\bf v}}+M^{-1}\frac{\partial}{\partial {\bf v}}\left[{\bf v}\beta(r,v)f\right]=0
\end{equation}
\end{widetext}
for the atomic phase space density $f({\bf r},{\bf v},t)$, where
\begin{equation} \label{e7}
{\bf F}_{\rm mf}({\bf r})=\int {\bf F}_{\rm rsc}({\bf r}^{\prime},{\bf r})f({\bf r}^{\prime},{\bf v})d{\bf v}d{\bf r}
\end{equation}
and $M$ is the mass of the atoms. Heating by spontaneous emission and photon exchange \cite{Bur,Ell94} has been neglected, since for the 
densities considered in this work 
the corresponding thermal pressure is much smaller than the pressure resulting from the effective atomic repulsion.

Note that eq.(\ref{e7}) goes beyond a local-density approximation \cite{Ell94}, retaining the complete position dependence of $\sigma_{\rm rsc}$ and the density dependence of ${\bf F}_{\rm rsc}$.
In fact, this nonlocal space dependence of all forces in eq.(\ref{e6}) in addition to their local dependence on the atom position renders a direct numerical solution of eq.(\ref{e6}) very demanding. Alternatively, we apply an efficient numerical procedure based on a test-particle treatment, similar to particle-in-cell methods \cite{Bir91},  frequently used for plasma physics problems. More specifically, we represent the atomic density by an ensemble of $N_{\rm t}<10^6$ test particles, whose number is typically chosen to be less than the actual particle number to reduce the numerical effort. The respective absorption cross sections and masses of the test particles are adjusted, such that the results are independent of the number $N_{\rm t}$ of test particles. By propagating every particle according to the forces eq.(\ref{e3}) and (\ref{e7}) we obtain the time dependent density from which we calculate the local intensities and the resulting forces to advance the next timestep.
\begin{figure}[b]
\begin{center}
\resizebox*{0.29\textwidth}{!}{\includegraphics{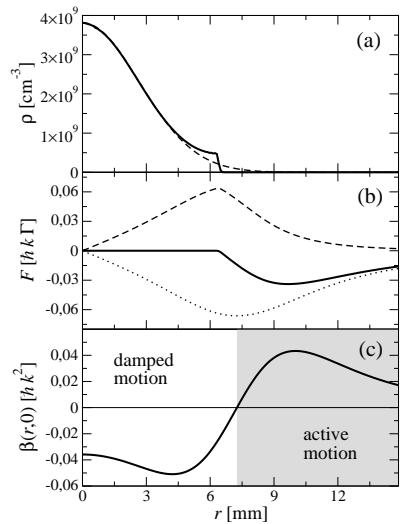}}
\caption{(a) Radial density profile (solid line) of trapped Rubidium atoms with $\delta=-1.5\Gamma$, $I=1.0$mW/cm and $\Gamma/\mu=4.7$mm together with a Gaussian fit (dashed line). (b) Radial dependence of $F_{\rm mf}$ (dashed line), $F_{\rm trap}$ (dotted line) and $F_{\rm mf}+F_{\rm trap}$ (solid line). (c) Radial dependence of the damping constant $\beta(r,v=0)$. The gray shaded area marks the region of active atomic motion.}
\label{fig1}
\end{center}
\end{figure}

To study its stationary properties we evolve the atomic cloud until it relaxes to the selfconsistent, stationary solution of eq.(\ref{e6}), 
which we found to exist only below a critical atom number $N_{\rm c}$. Fig.\ref{fig1}a shows the calculated stationary density profile 
for $N=1.15\times10^9$ Rubidium atoms and typical MOT-parameters of $I=1.0$mW/cm$^2$, $\delta=-1.5\Gamma$ and $\Gamma/\mu=4.7$mm 
(corresponding to $9$G/cm)\cite{Lab05}. As can be seen, the calculated density is well described by a truncated Gaussian profile. 
As the atom number is decreased the truncation radius $R$ decreases relative to the rms-width of the corresponding Gaussian, 
ultimately  leading to a transition into a uniform density profile.
Similar changes in the density profile have also been reported in MOTs, where the nonlinearity of the potential arises from sub-Doppler trapping mechanisms \cite{Tow95,Jun99}. In the present case the observed transition results from the nonlinearity and the position dependence of the reabsorption cross section and, hence, can not be found under the assumption of linear trapping forces and pure Coulomb-like interactions \cite{Ses91,Ste92,Arn00}.

\begin{figure}[t]
\begin{center}
\resizebox*{0.71\columnwidth}{!}{\includegraphics{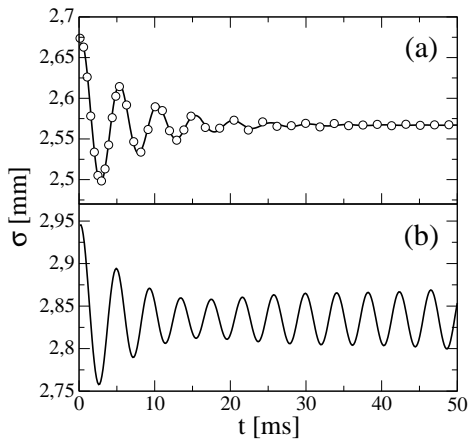}}
\caption{Relaxation of the MOT's RMS-radius after switching the detuning from $\delta_0=-1.55\Gamma$ to $\delta=-1.5\Gamma$ for two different particle numbers of $N=9\times10^8$ (a) and $N=1.3\times10^9$ (b). The remaining  parameters are the same as in fig.\ref{fig1}. The solid line in (a) shows a fit of $\Delta\sigma\sin(\omega t+\phi)e^{-t/\tau}+\sigma_{\infty}$ to the numerical data (circles). The solid line in (b) shows the numerical data in the oscillating regime of $N>N_{\rm c}$.}
\label{fig2}
\end{center}
\end{figure}

\begin{figure}[t]
\begin{center}
\resizebox*{0.64\columnwidth}{!}{\includegraphics{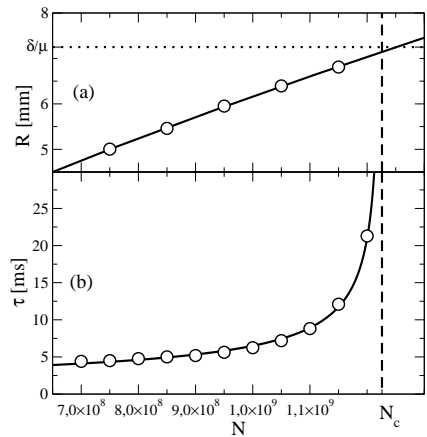}}
\caption{(a) Calculated MOT size as a function of the atom number (circles) fitted by a power-law dependence (solid line). The dotted line corresponds to the critical radius $R_{\rm c}=\delta/\mu$ and the dashed line in both figures indicates the critical atom number $N_{\rm c}$ beyond which the dynamical instability sets in. (b) Real part of the MOT's stability coefficient (circles) fitted by $\tau\propto \left(N-N_{\rm c}\right)^{\kappa}$ with $N_{\rm c}=1.226\times10^9$ and $\kappa=0.55$.}
\label{fig3}
\end{center}
\end{figure}

Let us now turn to the most striking result of our calculations.
As we further increase 
the number $N$ of atoms the cloud becomes unstable at a critical atom number $N_{\rm c}$, corresponding to a critical 
radius $R_{\rm c}$. By varying the various MOT-parameters, we find that the critical radius is uniquely determined by the 
relation $R_{\rm c}=\delta/\mu$ (see fig.\ref{fig3}a), confirming the conclusion reached in \cite{Lab05}. This fact is illustrated 
in fig.\ref{fig1}b and \ref{fig1}c, where we show the radial 
dependence of the trapping and interaction force as well as the 
damping 
constant $\beta(r,v=0)$. The damping constant $\beta(r,0)$ reverses 
its sign at $R_{\rm c}=\delta/\mu$. Hence, any small velocity of atoms 
outside of $R_{\rm c}$ will be enhanced. While inward moving particles 
will be damped again when entering the negative-$\beta$ region, outward 
moving atoms around $R_{\rm c}$ are further accelerated away from the trap center, since the single atom light pressure force is largely balanced by the interaction force around $r=R_{\rm c}$. 
Their motion around the fixed point ($r=R_{\rm c}, v=0$) will become unstable and limited by the non linear terms of the force. At larger distances, however, the total force reverses its sign again, since the interaction force decreases much more rapidly than the trapping, due to the radially increasing Zeeman shift (see fig. \ref{fig1}b).
Hence, if during the expansion, the atoms did not acquire a velocity beyond the capture range of the MOT, a stable limit cycle will be reached. 

In order to characterize the onset of the instability we analyze the clouds RMS radius $\sigma=\sqrt{\left<r^2\right>/3}$ and study 
its sensitivity against a small perturbation. More precisely, we start from a stationary density corresponding to some 
detuning $\delta_0$ which is instantly increased to $\delta$ (closer to resonance), leading to damped oscillations of $\sigma$ towards its new 
equilibrium value $\sigma_{\infty}$, as shown in fig.\ref{fig2}a. From a fit to a damped harmonic 
oscillation $\sigma=\Delta\sigma e^{-t/\tau}\sin(\omega t+\phi)+\sigma_{\infty}$ we obtain the damping 
time $\tau$ and frequency $\omega$ corresponding to the real and imaginary part of the respective 
Lyapunov exponent $\lambda=\tau^{-1}+i\omega$ (fig.\ref{fig3}b). For increasing $N$ and the 
parameters of fig.\ref{fig2} the instability sets in at an atom number of $N_{\rm c}=1.226\times10^9$ and with a 
critical exponent of $0.55$. On the other hand, the frequency of the cloud oscillation evolves continuously through 
the instability threshold (see fig.\ref{fig2}), indicating that the onset of the instability proceeds via an 
supercritical Hopf-bifurcation.

\begin{figure}[t]
\begin{center}
\resizebox*{0.83\columnwidth}{!}{\includegraphics{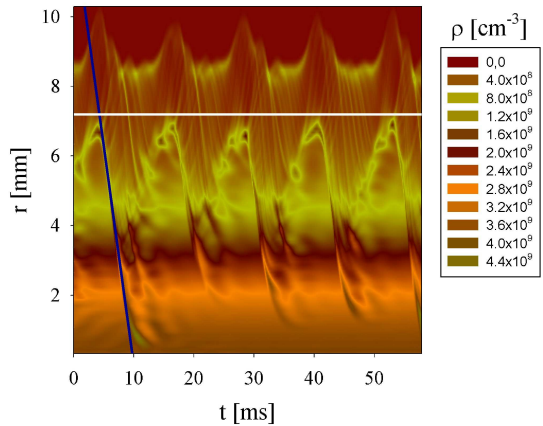}}
\caption{(color online) Spatio-temporal evolution of the atomic density for $N=1.7\times10^9$. The MOT parameters are the same as in fig.\ref{fig1}. The lines are discussed in the text.}
\label{fig4}
\end{center}
\end{figure}

A reduction of the system properties to a single quantity like the cloud's RMS-radius is clearly helpful for understanding 
the transition into the oscillating regime. On the other hand the fully resolved space-time evolution of the 
atomic density such as shown in fig.\ref{fig4} reveals much more detailed information about the complicated dynamics of the cloud. 
Indeed, the complex density patterns at larger atom numbers (see fig.\ref{fig4}) shows that the oscillation dynamics is much 
more complex than a simple breathing mode, as suggested by the simple size oscillations close to the instability threshold (see fig.\ref{fig2}b).

In fact, the oscillation is triggered by an outer fraction of atoms, which gain energy as they move 
in and out of the active region of $r>R_{\rm c}$, which is indicated by the horizontal white line in fig.\ref{fig4}. When 
bouncing back on the low-energetic atoms, the gained energy is deposited by exciting a density wave just inside the region 
with $\beta(r,0)<0$. Subsequently, the formed nonlinear excitation propagates towards the trap center along the diagonal 
blue line drawn in fig.\ref{fig4} and thereby loosing energy, mostly due to the damping by the cooling lasers. As can be 
seen in fig.{\ref{fig4}} this not only leads to a flattening and broadening of the density wave until it disappears, but 
also to a deceleration as indicated by the deviation of the moving maximum from the blue line at smaller distances. At the 
same time, the edge region of the atomic cloud starts to relax, causing some atoms to be again accelerated away from the 
center and the whole process repeats itself. Although this scenario clearly provides the basic mechanism for the 
observed oscillations, our calculations reveal a number of finer details (see fig.\ref{fig4}) still to be understood. 
Moreover, additional damping mechanisms, similar to Landau-damping of plasma waves might also
play a role for the system dynamics, raising the interesting question of how the present nonlocal 
position dependence of the effective charges manifests itself in known plasma kinetic effects.

In conclusion, large clouds of magneto-optical confined atoms have been 
found to exhibit a very complex nonlinear dynamics. Our theoretical description has revealed the onset of a deterministic instability connected with self-sustained oscillations in agreement with recent experiments \cite{Lab05}.
It has been found that a number of different effects, such as the attenuation of the trap lasers, rescattering of the absorbed laser light as well as the position dependence of the respective absorption cross sections are {\emph all} necessary to explain the observed phenomenon. A stability analysis of the MOT size has shown that the transition proceeds via a supercritical Hopf-bifurcation. The obtained density evolution revealed the build-up of complex nonlinear excitations driven by the combined action of the light-pressure force and the effective atomic interaction, which results in an active atomic motion at large distances. Similar types of active or self-driven motion are currently discussed in a broad range of different applications, such as collective swarm dynamics \cite{Erd05}, propagation of waves \cite{Sak02} or dissipative solitons \cite{Or98} in reaction-diffusion systems or grain motion in dusty plasmas \cite{Tri03}. Hence, we believe that large clouds of magneto-optical confined atoms provide an ideal laboratory system for further exploration of the rich spectrum of self-driven motion, including variable system geometries, effects of external driving and possibilities to control the system dynamics.

TP would like to thank for the kind hospitality during a stay at the Institut Non Lin\'eaire de Nice where major parts of the work have been performed and acknowledges support from the ESF through the Short Visit Grant 595 and from the NSF through a grant for the Institute of Theoretical Atomic, Molecular and Optical Physics (ITAMP) at Harvard University and Smithsonian Astrophysical Observatory.

\end{document}